\documentstyle[12pt]{article}
\begin{document}
\baselineskip 24pt
\vspace{0.5 truecm}

\begin{center}
{\Large \bf  Isotopic and Microcanonical Temperatures in  
Nuclear Multifragmentation}
\end{center}
\vspace{0.5 truecm}
\begin{center}
{\large J.P. Bondorf$^{1}$, A.S. Botvina$^{2,3}$ 
and I.N. Mishustin$^{1,4}$}
\end{center}
\vspace{0.5 truecm}

{\it 
$^{1}$Niels Bohr Institute, DK-2100 Copenhagen {\O}, Denmark

$^{2}$Dipartimento di Fisica and INFN, 40126 Bologna, Italy

$^{3}$Institute for Nuclear Research, Russian Academy of Science, 
 117312 Moscow, Russia

$^{4}$Kurchatov Institute, Russian Scientific Center, 123182 Moscow, Russia}

\vspace{0.5 truecm}
\begin{abstract}
A systematic comparison of different isotopic temperatures with the 
thermodynamical temperature of a multifragment system is made on the basis 
of the Statistical Multifragmentation Model. It is demonstrated that isotopic 
temperatures are strongly affected by the secondary decays of hot primary 
fragments and the population of particle-stable excited states in final 
fragments. The He-Li temperatures, measured recently by the ALADIN group, 
are reproduced fairly well both as a function of excitation energy and 
bound charge. Our analysis confirms the anomaly in the nuclear caloric curve. 
\end{abstract}
\vspace{0.3 truecm}

\hspace{0.5 truecm}PACS numbers: 25.70-z, 25.70Pq, 24.60-k
\vspace{0.5 truecm}

Nuclear multifragmentation in heavy-ion reactions is intensively studied 
at present both theoretically and experimentally. One of the main goals is to 
investigate properties of nuclear matter away from the ground state. A most 
interesting question here is how multifragmentation is related to a liquid-gas 
phase transition in a finite nuclear system. To answer this question one needs 
observables which bring information about the thermodynamical state of the 
system, in particular, its excitation energy and temperature.
Then a phase transition should manifest itself by an anomaly in 
the caloric curve, i.e. temperature as a function of excitation energy. 
According to the statistical model prediction \cite{bondorf}, the nuclear 
caloric curve behaves like in an ordinary liquid-gas phase transition: 
initially the temperature increases, at excitation energies between 3 and 
10 MeV/nucleon it stays almost constant at about 5--6 MeV, 
and then grows again. 
The first regime corresponds to the compound nucleus (liquid phase), the 
second one, to the multi--fragment mixture (coexistence phase), and the third 
one, to an assembly of nucleons and lightest clusters (gaseous phase).

The first measurements of the nuclear caloric curve have been made 
only recently by the ALADIN group \cite{pochod}. They indeed revealed  
an anomalous behaviour of the nuclear caloric curve similar to that predicted
by the statistical model \cite {bondorf}. In the experiment the so-called 
isotopic temperature, $T_{isot}$, was extracted from the double ratio of 
helium and lithium isotope yields. At present nuclear temperature 
measurements are in fast progress. Several groups have reported results 
on nuclear caloric curves for different reactions and with different 
isotope thermometers \cite{tsang,ma,wada,xi}.
Therefore, it is very important now to understand how these isotopic 
temperatures are related to the thermodynamical temperatures of excited
nuclear systems at the stage of their break-up.

According to the method suggested by Albergo et al. \cite{albergo}, the 
isotopic temperature is expressed through the double ratio of 
isotope yields as 
\begin{equation} \label{tem}
T_{isot} = \frac{B} {\ln(a \cdot R)}~~.
\end{equation} 
Here $R=(Y_1/Y_2)/(Y_3/Y_4)$, $B=(B_1-B_2)-(B_3-B_4)$. $Y_i$ and $B_i$ are 
the $i$-th isotope yield and binding energy, $a$ is a constant determined by 
spin degeneracy factors and masses of the isotopes. 
The indexes $i=$1, 2, 3 and 4 refer to the isotopes with masses and 
charges (A,Z), (A+1,Z), (A$^{'}$,Z$^{'}$) and (A$^{'}$+1,Z$^{'}$),
respectively. 

It is clear that this expression corresponds to the 
grand canonical approximation assuming thermal and chemical equilibrium.
Moreover, it is assumed that all fragments are 
produced simultaneously at the same $T$ and only in their ground states. 
These assumptions are too crude for finite and highly excited nuclear systems 
under consideration. A more realistic approach should include at least two
important modifications: first, the microcanonical treatment of the break-up 
channels i.e. taking into account exact conservation laws for baryon number, 
charge and energy, and second, the feeding of isotope yields from the 
de-excitation of hot primary fragments after the break-up. 
The importance of secondary decays was demonstrated earlier by several authors
(see e.g. refs. \cite{kolomiets,huang,gulminelli}). 
Statistical models of multifragmentation (see reviews \cite{gross,PR})
provide a natural framework for introducing these modifications. These models 
are very successful in describing many observed characteristics of nuclear 
multifragmentation (see examples in refs. \cite{botvina1,dagostino}).

The calculations below are made within the standard version of the Statistical 
Multifragmentation Model (SMM) which was used for the first calculation of the 
nuclear caloric curve \cite{bondorf}. 
Here we outline only some general features of the model 
(see details in ref. \cite{PR}). It is assumed that at the break-up time the 
system consists of primary hot 
fragments and nucleons in thermal equilibrium. Each break-up channel, $f$, is 
specified by the multiplicities of different species, $N_{AZ}$, which are 
constrained by the total baryon number $A_0$ and charge $Z_0$ of the system. 
The probabilities of different break-up channels are calculated 
in an approximate microcanonical way according to their statistical weights,
$W_f \propto \exp\left[S_f(E^*,V,A_0,Z_0)\right]$, where $S_f$ is the entropy
of a channel $f$ at excitation energy $E^*$ and break-up volume $V$. 

Translational degrees of freedom of individual fragments are described by 
the Boltzmann statistics while the internal excitations of fragments with 
$A>4$ are calculated within the liquid-drop model with Fermi-gas
level density.  
An ensemble of microscopic states corresponding to the break-up channel $f$ 
is characterized by a temperature $T_f$ which is 
determined from the energy balance equation
\begin{equation} \label{ene}
\frac{3}{2}T(m-1)+\sum_{(A,Z)}E_{AZ}(T)N_{AZ}+E_f^C(V)-Q_f=E^*~~.
\end{equation} 
Here $m=\sum N_{AZ}$ is the total fragment multiplicity, first term comes 
from the translational motion, second term includes
internal excitation energies of individual fragments and third term is
the Coulomb interaction energy, $Q_f$ is the $Q$--value of the channel $f$.
The excitation energy $E^*$ is measured with 
respect to the ground state of the compound nucleus ($A_0$,$Z_0$).
In our semi-microcanonical treatment $E^*$ is fixed for all fragmentation 
channels while the temperature $T_f$ fluctuates from channel to channel.

The total break-up volume is parametrized as $V=(1+\kappa) V_0$,
where $V_0$ is the compound nucleus volume at normal density and the 
model parameter $\kappa$ is the same for all channels. The choice of 
$\kappa$ is motivated by the requirements: a) to avoid overlaps between the 
fragments and b) to provide a sufficient reduction of the Coulomb barrier, 
as seen in the kinetic energy spectra. The 
entropy associated with the translational motion of fragments is determined 
by the  ``free'' volume, $V_f$, which incorporates the excluded volume effects. 
In general $V_f$ depends on the break-up channel and therefore cannot be 
fixed to a constant, $\kappa V_0$, as often assumed. In the SMM we parametrize 
$V_f(m)$ in such a way that it grows almost linearly with the primary 
fragment multiplicity $m$ or, equivalently, with the excitation energy 
$\varepsilon^*=E^*/A_0$ of the system \cite{PR}. According to this 
parametrization, $V_f(m)$ vanishes for the compound nucleus ($m=1$) and 
increases to about 2$V_0$ at $\varepsilon^*\approx$ 10 MeV/nucleon.

At given inputs $A_0$, $Z_0$ and $\varepsilon^*$ the individual 
multi-fragment configurations are generated by the Monte Carlo method. 
After the break-up hot primary fragments loose their excitation.
The most important de-excitation mechanisms included in the SMM \cite{PR} 
are the simultaneous Fermi break-up of lighter fragments  ($A\leq 16$) 
and the evaporation from heavier fragments, including the compound-like  
residues. In this respect SMM essentially differs from the 
QSM type models \cite{gulminelli} where the compound-like channels 
are completely ignored (see discussion in ref. \cite{botvina2}).

Now we turn to numerical simulations of the multifragmentation on the basis 
of SMM. First of all  we present results for a well defined source i.e. 
an excited $^{197}$Au nucleus. The caloric curve is calculated by first 
solving eq. (2) for each particular channel and then averaging $T_f$ over a 
large number of break-up channels. In Fig.~1 (top) different curves correspond 
to different choices of volume parameters. If the standard parametrization 
$V_f(m)$ is used, the caloric curve is quite flat in the $\varepsilon^*$ 
region between 3 and 10 MeV/nucleon. This is a signature of a large heat 
capacity in the transition region. Even a backbending is possible if 
the total volume $V$ is not very large, say only 3$V_0$ ($\kappa$=2). 
In contrast, if the free volume would be fixed to $V_f=\kappa V_0$ for all 
channels, the temperature would increase gradually with $\varepsilon^*$. 
Nevertheless, as seen in Fig. 1, some flattening in the caloric curve is 
predicted also in this case.  
The reason for the different behaviour is clear: at $\varepsilon^*<10$ 
MeV/nucleon the multiplicity-dependent free volume $V_f(m)$ is smaller than 
$\kappa V_0$ that leads to a higher temperature of the system. 
In the following calculations we use $\kappa$=2 and the standard SMM
parametrization of $V_f(m)$, which gives a plateau in the caloric curve.
As we will see below such an behaviour is favoured by the data.

The characteristics of the system change drastically when $\varepsilon^*$ 
increases from 3 to 10 MeV/nucleon. In the lower part
of Fig.~1 we display several observables calculated after the completion 
of all secondary decays. A heavy residue, usual for the evaporation-like 
processes, practically disappears. This is signalled by the maximum fragment 
charge, $Z_{max}$, which drops rapidly from 60 to about 6 in this region.   
At the same time, the number of Intermediate Mass Fragments (IMFs: 
$3\leq Z\leq 20$)  
first increases and then goes through the maximum, $N_{imf}\approx$ 8, in 
the end of this region. The multiplicity of all charged particles, $N_{ch}$,
grows with $\varepsilon^*$ almost linearly. It is interesting to note that 
in the transition region the number of free neutrons, $N_{neu}$, is almost 
constant and close to the neutron excess in the initial $^{197}$Au nucleus.
This happens because the system breaks up predominantly into fragments with 
$N\approx Z$ (see also \cite{botvina2}). By comparing upper and lower parts 
of Fig.~1 one can conclude that the neutron 
multiplicity is nearly proportional to the temperature of the system but not 
to the excitation energy. 

We have also calculated final isotope yields in
the disintegration of $^{197}$Au nucleus. Several isotopic temperatures 
were obtained by applying formula (1) to different isotope pairs. Results are 
shown in Fig.~2 together with the microcanonical temperature $T_{micr}$. 
One can see that the plateau is almost washed out and all isotopic temperatures
increase gradually with $\varepsilon^*$. 
This behaviour can be explained by the de-excitation of hot primary fragments
leading to their cooling and side-feeding of isotope yields.
Since the energy conservation is controlled at all stages of the 
calculations, the SMM leads naturally to the cooling of emitters in 
endothermic processes responsible for the fragment de-excitation. 
In the case of sequential evaporation
the first fragments are emitted from a source characterized 
by the emission temperature $T_{micr}$. 
But the next generation of fragments comes from a cooler residue 
leading to a lower apparent temperature \cite{wada}. This cooling 
mechanism can explain partly 
the difference between the isotopic temperatures and $T_{micr}$ 
at lower excitation energies ($\varepsilon^*=1\div 6$ MeV/nucleon), when 
heavy residues ($Z_{max}>20$) survive in the break-up. 
At $\epsilon^*\geq 3$ MeV/nucleon another de-excitation mechanism 
becomes increasingly important, i.e. the one-step Fermi break-up where
only particle-stable decay products are allowed. 
It is mainly responsible for the production of light isotopes, in particular 
He and Li, through the deep disintegration of heavier fragments ($A\sim15$). 
Since the available energy (per nucleon) is considerably lower in this process 
than in the primary break-up, the apparent isotopic temperatures are also lower.
Finally, at high excitation energies, $\epsilon^*\geq 10$ Mev/nucleon, when 
predominantly light fragments are formed,  
the reduction of the available energy for secondary break-up becomes less 
important and isotopic temperatures, in average, approach $T_{micr}$.

From  Fig.~2 one can also see that the temperature measurements can 
be significantly obscured by the irregularities in the excited states of light 
fragments. In our standard calculations the final isotope yields include the 
fragments in particle-stable ground and excited states decaying by the 
$\gamma$-emission. For the considered isotopes they are: 
3.56 MeV for $^{6}$Li, 0.48 MeV for $^{7}$Li, 0.43 MeV for $^{8}$Li, 
3.37, 5.96, 6.18 and 6.26 MeV for $^{10}$Be. No such excited states are seen
in $^{3,4}$He and $^9$Be and therefore only ground states are included for 
these  nuclei.
One can see that the deviations from the true temperature are especially 
large (curve $a$) in the case when one of the isotopes has 
many and the other, only a few or no excited states, like e.g. in the 
$^{10}$Be--$^{9}$Be pair. If the excited states in $^{10}$Be are 
artificially switched off, the corresponding isotopic temperature (curve $b$) 
changes drastically and follows the common trend. To suppress the fluctuations
associated with the nuclear structure effects one can use an ensemble of 
isotope thermometers as suggested in ref. \cite{tsang}.
Another possibility is to use isotope pairs with only a few low-lying states 
which almost compensate each other, like e.g. $^{7}$Li-$^{8}$Li.  
In this respect it is preferable to use thermometers
with isotopes not heavier than lithium, such as the He-Li one.
But in this case one is facing another problem, i.e. the contamination of yields
by the pre-equilibrium emission prior to the break-up. This 
contribution is most important for lighter fragments and can be evaluated only 
on the basis of dynamical simulations. 

To apply SMM for analyzing experimental data one needs to know the 
characteristics (masses, charges, excitation energies) of thermalized emitting 
sources. A clear identification of such sources is made only in a few 
cases. One example is given in ref. \cite{dagostino} where the emitting 
source with mass 315, charge 126 and thermal excitation energy of about 5 
MeV/nucleon was found for central Au+Au collisions at 35 AMeV. The SMM 
calculations reproduce nicely the fragment charge distribution for this 
reaction yielding the emission temperature of about 6 MeV\cite{dagostino}. 
By inspecting Fig.~2 one can see that there is no contradiction 
between the value $T$($^{6,7}$Li/$^{3,4}$He)$\approx$ 4.6 MeV measured 
for this reaction 
\cite{huang} and the SMM prediction of $T_{micr}\approx$ 6 MeV. In accordance
with experiment is also that the Be-Li temperature is much higher than other
isotopic temperatures. On the other hand, two other isotopic temperatures
presented in Fig.~2, $T$($^{2,3}$H/$^{3,4}$He) and $T$($^{7,8}$Li/$^{3,4}$He), 
are predicted too high and in inverse order compared to 
$T$($^{6,7}$Li/$^{3,4}$He)
\cite{huang}. Our analysis shows that the yields of neutron-rich isotopes, 
such as $^3$H and $^8$Li, are quite sensitive to the $N/Z$ ratio in the 
decaying thermalozed source. The results of Fig.~2 correspond to the $197$Au 
nucleus with $N/Z\approx$ 1.5. The correct ordering of the isotopic 
temperatures can be achieved by adjusting the $N/Z$ ratio in the source. 

Finally we present our analysis of the ALADIN data \cite{pochod,imme} 
for peripheral Au+Au collisions at 0.6 and 1 AGeV (see also \cite{wada}). In 
these experiments only  
fragments from the projectile spectators were detected. Therefore, here  
we are dealing with a wide ensemble of emitting sources
produced at different impact parameters. As known \cite{botmis}, the 
masses and excitation energies of these sources are strongly affected 
by the pre-equilibrium emission. 
Nevertheless, the ensemble 
of thermalized sources can be reconstructed by backtracing the measured
characteristics of produced fragments \cite{botvina1,deses}.   

In the SMM calculations presented in Fig.~3 we have considered two 
different ensembles of emitting sources obtained in refs. \cite{pochod} and 
\cite{botvina1}. The ``experimental'' 
ensemble of ref. \cite{pochod} has a wider distribution
in excitation energy (up to about $\varepsilon^*\approx$ 14 MeV/nucleon) than 
the ``theoretical'' ensemble of ref. \cite{botvina1} which is 
limited at $\varepsilon^*\approx$ 8 MeV/nucleon. As seen from Fig.~3 the 
observed He-Li temperatures are better reproduced by the experimental 
ensemble. But this ensemble is certainly contaminated by the early 
emitted H and He fragments which were not separated in the data analysis.
Obviously their admixture is larger at higher excitation energies.
On the contrary, 
in ref. \cite{botvina1} the sources were reconstructed by using 
the characteristics of fragments with $Z\geq$3 which are less affected by the 
pre-equilibrium emission. Therefore, we expect that after separating 
early emitted H and He fragments experimental points will shift closer to the 
prediction of the theoretical ensemble.

For both ensembles we get a more steep
increase of the isotopic temperature with excitation energy than the 
experimental data show (Fig.~3, top). Also within the present version of SMM we 
cannot reproduce the low temperatures extracted from the relative level 
population in light fragments such as $^5$Li. 
Our preliminary calculations show that the agreement with experiment
can be improved by reducing excitation energies of primary fragments and thus
suppressing their secondary decay contribution. This and other modifications of
the model are under investigation now.

In conclusion, on the basis of SMM we have demonstrated that 
the secondary de-excitation processes and irregularities in  
particle-stable excited states of fragments may cause significant 
deviations of isotopic temperatures from the thermodynamical temperature 
of the decaying system. Our analysis shows that the ALADIN data are 
consistent with the anomaly in the nuclear caloric curve.
For future studies of the nuclear caloric curve it is very 
important to separate the contribution of light clusters emitted at early 
non-equilibrium stages of the reaction. Therefore, the determination of 
the temperature and excitation energy should be accompanied by a thorough 
kinematical analysis of emitting sources and fragment spectra. 

\vspace{0.5 truecm}
We thank D.H.E. Gross, W. Friedman, W. Lynch, J. Pochodzalla, U. Schr\"oder, 
W. Trautmann and B. Tsang for useful discussions. A.S. Botvina thanks 
Istituto Nazionale di Fisica Nucleare (Italy) and the Niels Bohr 
Institute for hospitality and support, I.N.Mishustin thanks 
the Carlsberg Foundation (Denmark) for financial support.
This work was supported in part by the EU-INTAS grant No. 94-3405.

\vspace{0.5 truecm}
{\large \bf  {Figure captions}}\\

{\bf Fig.1:} {Top: Caloric curves as predicted by the SMM simulations  
for an excited $^{197}$Au nucleus. Results are shown for four different
choices of volume parameters characterizing the break-up 
configuration (see the text). 
Bottom: Some observable characteristics as functions of excitation energy 
in multifragmentation of $^{197}$Au nucleus after de-excitation of primary 
fragments. Z$_{max}$ is the maximum fragment charge, $N_{imf}$ is the 
multiplicity of intermediate mass fragments ($3\leq Z\leq 20$), 
$N_{ch}$ and $N_{neu}$ are the total numbers of charged particles and free 
neutrons.}
\vspace{0.5 cm}

{\bf Fig.2:} {Isotopic temperatures for four isotope pairs (indicated 
in the figure) versus excitation energy calculated for $^{197}$Au 
by applying formula (1) to final isotope yields. The microcanonical 
temperature of the decaying nucleus is the solid line.
\vspace{0.5 cm}

{\bf Fig.3:} {He--Li temperatures (scaled by factor 1.2) versus excitation 
energy $\varepsilon^*$ (top) and bound charge $Z_{bound}$ (bottom) for 
projectile spectators produced in Au + Au collisions at 0.6 and 1.0 AGeV. 
Symbols represent the ALADIN data for 0.6 AGeV \cite{pochod,imme} (dots) 
and 1.0 AGeV \cite{imme} (triangles). The SMM calculations are made for 
two ensembles of thermalized sources: 1 from ref. \cite{botvina1} and 
2 from ref. \cite{pochod}.}

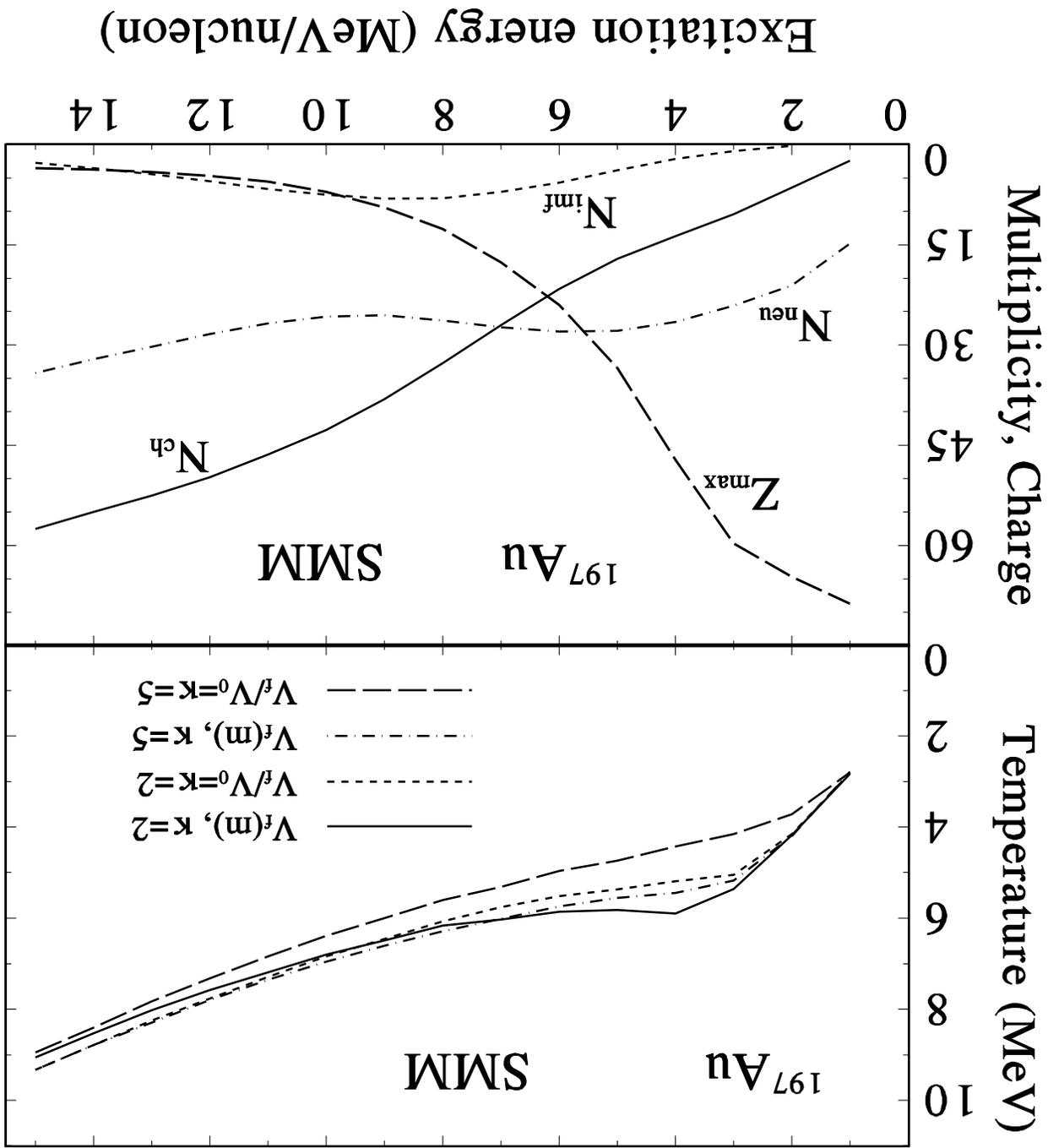
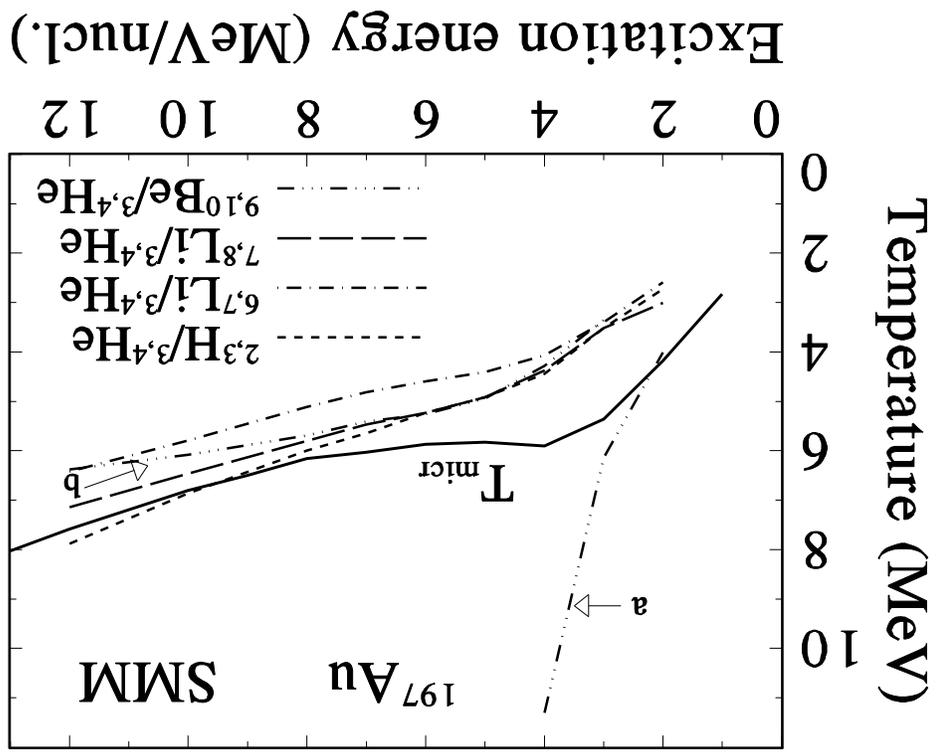
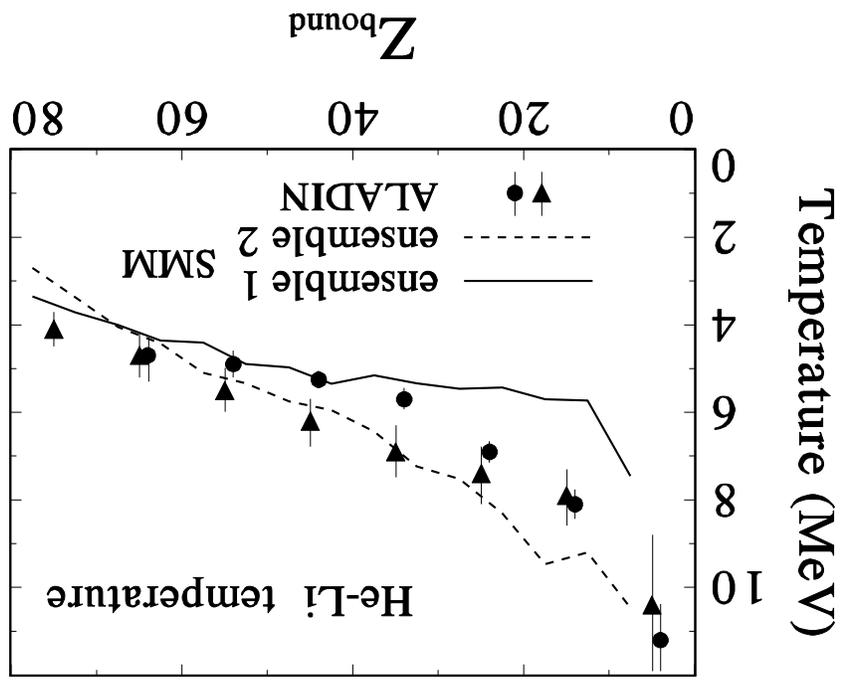
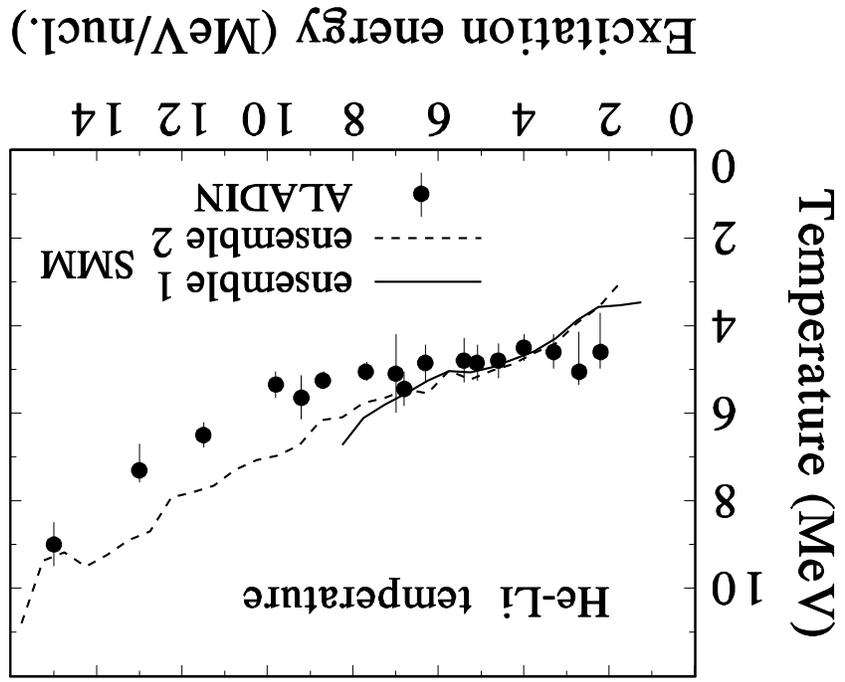
\end{document}